\documentclass{article}
\usepackage{epsf}
\usepackage[dvips]{graphicx}
\begin{document}
\baselineskip7mm
\title{Certain aspects of regularity in scalar field 
cosmological dynamics}
\author {A.Toporensky and P.Tretyakov}
\date{}
\maketitle
\hspace{8mm} {\em Sternberg Astronomical Institute,
Universitetsky prospekt, 13, Moscow 119992, Russia}

\begin{abstract}
We consider dynamics of the FRW Universe with a scalar field. Using
Maupertuis principle we find a curvature of geodesics flow and show that
zones of positive curvature exist for all considered types of scalar field
potential. Usually, phase space of systems with the positive curvature
contains islands of regular motion. We find these islands numerically 
for shallow scalar field potentials. It is shown also that beyond the physical
domain the islands of regularity exist for quadratic potentials as well. 
 
\end{abstract}

\section{Introduction}
It is known from 70-th of the last century that contraction phase of a closed
Universe filled with a massive scalar field can be followed by expansion for
some particular initial condition set \cite{P-F, S}. On the other hand, every expansion stage of such
Universe is ultimately followed by a contraction one. These two features of dynamics
(which are specific for a closed Universe in contrast to open or flat worlds) result in
a rather complicated behavior which in some situations can be chaotic.

A chaotic dynamics in massive scalar field cosmology was first found by D. Page in
\cite{Page}, and have been studied in detail in \cite{Shellard, Levin} with the corresponding
discussion on the meaning of chaos in General Relativity. The chaotic dynamics in question
represents an example of a transient chaos with a structure of "chaotic
repellor" -- a countable set
of unstable periodic and
uncountable set of unstable aperiodic trajectories escaping cosmological singularity.
Both sets have zero measure in initial condition space. An useful toy model of such kind
of dynamics (elastic scattering on three discs on a plane) have been described in \cite{Rice}.

All these early results have been found for a massive scalar field -- scalar field
with the potential in the form $V(\phi)=m^2 \phi^2/2$, where $m$ is the scalar field mass.
Studies of other forms of the potential reveal several 
different form of dynamics with  transitions from one form to another (for a short review see \cite{mysigma}). 

The equations of motion can be derived from the General Relativity action with a scalar field
\begin{equation}
S = \int d^{4} x \sqrt{-g}\left\{\frac{m_{P}^{2}}{16\pi} R +
\frac{1}{2} g^{\mu\nu}\partial_{\mu}\phi \partial_{\nu}\phi
-V(\phi)\right\},
\end{equation}
where $m_{P}$ is the constant parameter called the Planck mass, $R$ is the scalar curvature
of a space-time.
For a closed Friedman model with the metric
\begin{equation}
ds^{2} =  dt^{2} - a^{2}(t) d^{2} \Omega^{(3)},
\end{equation}
where
$a(t)$ is a cosmological scale factor, 
$d^{2} \Omega^{(3)}$ is the metric of a unit 3-sphere and
with homogeneous scalar field $\phi$
the action (1) gives the following equations:

\begin{equation}
\frac{m_{P}^{2}}{16 \pi}\left(\ddot{a} + \frac{\dot{a}^{2}}{2 a}
+ \frac{1}{2 a} \right)
+\frac{a \dot{\phi}^{2}}{8}
-\frac{a V(\phi)}{4} = 0,
\end{equation}
\begin{equation}
\ddot{\phi} + \frac{3 \dot{\phi} \dot{a}}{a}
+ V'(\phi) = 0.
\end{equation}
with two variables  - a scale factor $a$ and a scalar field $\phi$.

This system has one first integral of motion
\begin{equation}
-\frac{3}{8 \pi} m_{P}^{2} a (\dot{a}^{2} + 1)
+\frac{a^{3}}{2}\left(\dot{\phi}^{2} + 2 V(\phi)\right)  =
C.
\end{equation}

This integral plays the role of energy and is equal to zero
for cosmological solutions of the system (3)-(4). 

If $V(\phi)=0$, the system is integrable (in this case a scalar field has
the equation of state of stiff matter), though the dynamics is finite in time -- 
the Universe evolves from Big Bang singularity towards a Big Crunch singularity
through a point of maximal expansion. There is also another reason why we can
not develop a perturbation analysis for the system (3)--(5): it can be easily checked
that multiplying the scalar field potential $V$ by a constant with simultaneous 
rescaling of the scale factor $a$ and time $t$ does not change the system.
This means that there is no continuous limit when, for example, $m \to 0$
for the massive scalar field.  

An important property of the constraint equation (5) is 
that $\dot a^2$ and $\dot \phi^2$ enter in its left-hand side 
with opposite signs. Rewriting (5) in the form
\begin{equation}
-\frac{3 m_P^2}{8 \pi}\frac{\dot a^2}{a^2}+\frac{\dot \phi^2}{2}=\frac{3m_P^2}{8 \pi}\frac{1}{a^2}-V(\phi)
\end{equation}
it is easy to see that
\begin{itemize}
\item There are no forbidden regions in the configuration state.
\item The configuration state is divided to two regions -- the region
where the right-hand side of (6) is positive (and possible extrema
of the scale factor are located), and the region where it is negative
(where possible extrema of the scalar field are located).
\end{itemize}

The boundary between these two regions is the curve 
\begin{equation}
a^2=\frac{3}{8 \pi} \frac{m_{P}^2}{V(\phi)}.
\end{equation}

Extrema of the scale factor can exist only in the region where
$$
a^{2} \leq \frac{3} {8 \pi}  \frac{m_{P}^2}{V(\varphi)}.
$$
Using (3) it can be shown that
the possible points of
maximal expansion ($\dot a=0$, $\ddot a<0$) are localized inside the region
$$
a^{2} \leq \frac{1}{4 \pi} \frac{m_{P}^{2}}{V(\varphi)}
$$
while the possible points of transition from contraction to expansion 
(often called bounce) lie outside this
region being at the same time inside the region of scalar field extrema.

Zero-velocity points ($\dot a = \dot \phi =0$) of a trajectory, it they exist, should lie on the curve (7).
Numerical studies show that trajectories with these points play an
important role in the 
described chaotic structure. In particular, all primary (i.e.
having one bounce per period) periodical
trajectories have zero-velocity points as the points of bounce (see
numerical examples in \cite{Shellard}).

Numerical integrations show also that there are regions on the curve (7)
which can not contain
points of bounce. If a trajectory, starting from the curve (7) is
directed inside the zone of possible
extrema of the scale factor, it rapidly goes through a point of maximal
expansion and evolves
further towards a singularity. The condition for a trajectory to be
directed into the
opposite zone (the zone of
possible extrema of the scalar field) can be written as
\begin{equation}
\ddot \phi/\ddot a > d\phi(a)/da
\end{equation}
where the function $\phi(a)$ in the RHS is the equation of the curve
(7).
 
The case of equality in (8) corresponds to a trajectory, tangent to the
curve (7).
This situation was first described in \cite{Page}, and we call such point
as a Page point. For the system (1)-(3) the equation for the Page point
is \cite{we}
\begin{equation}
V(\phi_{page})=\sqrt{\frac{3 m_P^2}{16 \pi}}V'(\phi_{page})
\end{equation}

For power-law scalar field potentials the
condition (8) is satisfied if $\phi > \phi_{page}$,
and the corresponding
part of the curve (7) contains zero-velocity bounce points of periodical
trajectories. For exponential
potentials the condition (8) can be violated for all points on the curve
(7), and the whole chaotic
structure disappears \cite{we}. The dynamics in this case can be
called regular in the sense that the structure of chaotic repellor is absent,
and there are no trajectories escaping a  cosmological singularity. Global structure
of trajectories is the same as for $V=0$ case -- any trajectory starts and ends 
in a singularity. This does not mean that there are no local instabilities of trajectories
(see the next section), however, as the dynamics is finite in time these instabilities
are not important for the global picture.

In the next section we study local instabilities of trajectories, in Sec.3 we describe
a new regular regime for the system (3)--(5) , which appears when the scalar field potential
is sufficiently shallow.

\section{Local instability}

The system (3)-(4) can be considered as a Hamiltonian system
with 
\begin{equation}
{\cal H}= -\frac{2 \pi}{3 m_P^2 a} p_a^2 + \frac{1}{2 a^3} p_{\phi}^2 - \frac{3 m_P^2 a}{8 \pi} + V(\phi) a^3, 
\end{equation}
where the canonical variables are $q^i=(a, \phi).$

We rewrite (10) in the form
\begin{equation}
{\cal H}=a^{ij}p_i p_j + {\cal V}
\end{equation}
where 
$$a^{ij}=diag(-\frac{2}{3}\frac{\pi}{am^2_p},\frac{1}{2 a^3}),$$
$${\cal V} =  V(\phi) a^3 - 3 m_P^2 a/(8\pi).$$

According to Maupertuis principle we can introduce an auxiliary metric space with geodesics
coinciding with trajectories of initial system. This method of geodesics 
have been used actively since the middle of the last century \cite{Krylov}
in various physical problems from Yang-Mills 
fields \cite{Sav} to cosmology \cite{Gurzadyan2}.  
 Calculating  the Riemann
curvature ${\cal R}$ of this space we get the criterion of local
instability in the form ${\cal R} < 0$.  For the system
with a Hamiltonian in the form (11), the resulting curvature is 
\cite{Sav2,Rugh} 

$$
{\cal R}=\frac{(n-1)}{2(C-{\cal V})^3}\sum\left [(C-{\cal V})\frac{\partial
^2{\cal V}}{\partial q^i\partial q^j}a^{ij}-\frac{(n-6)}{4}\frac{\partial
{\cal V}}{\partial q^i}\frac{\partial {\cal V}}{\partial q^j}a^{ij} \right ],
$$
where $n$ is the number of degrees of freedom.

Taking $C=0$ we get the curvature for cosmological solutions
of our system: 

\begin{equation}
{\cal R}=\frac{-1}{2(\frac{3}{8}\frac{m_p^2}{\pi}a-a^3V)^3}\left [ a
V^{''}\left (\frac{3}{8}\frac{m_p^2}{\pi}-a^2V\right ) + a^3\left
(V^{'2}-4\frac{\pi}{m_p^2}V^2\right )
-\frac{3}{16}\frac{m_p^2}{\pi}\frac{1}{a} \right ],
\end{equation}
where $'$ denotes partial differentiating with respect to a scalar field $\phi$.

The curvature diverges at the curve (7), and this is a consequence of the fact that
the supermetrics $a^{ij}$ is not positively definite. This property is typical for
dynamical systems of General Relativity and discussed in \cite{Rugh}.

Three different cases are plotted in Fig.1. The first plot  (Fig.1(a)) represents the situation
of a massive scalar field with the potential $V=m^2 \phi^2/2$. We know that
a chaotic repellor exists in the corresponding dynamical system. The Fig.1(b)
is created for the potential $V=\cosh(\phi)-1$, for which it is known that the structure
of chaotic repellor is absent. 

We can see that in both cases there are zones with ${\cal R}>0$ as well as zones with
${\cal R} <0$. It means that studying only local properties of our system, it is impossible
to understand its global features. The chaotic system does not belong to the class 
of Anosov's U-systems (this would require ${\cal R} <0$ in all points of the configuration
space), which posses maximally strong statistical properties \cite{Anosov}.

\begin{figure}[h]
\includegraphics[scale=0.38, angle=0]{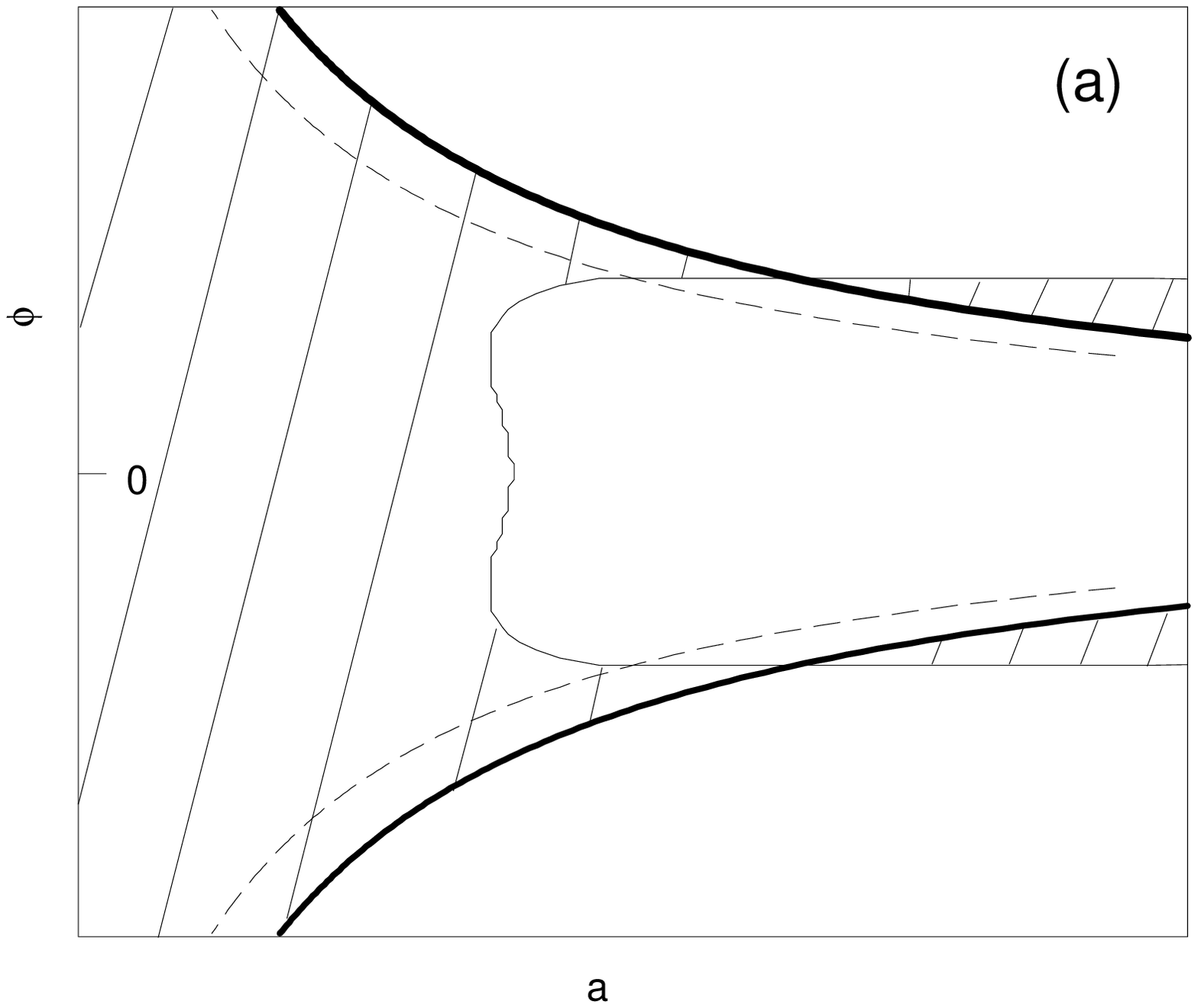}
\includegraphics[scale=0.38, angle=0]{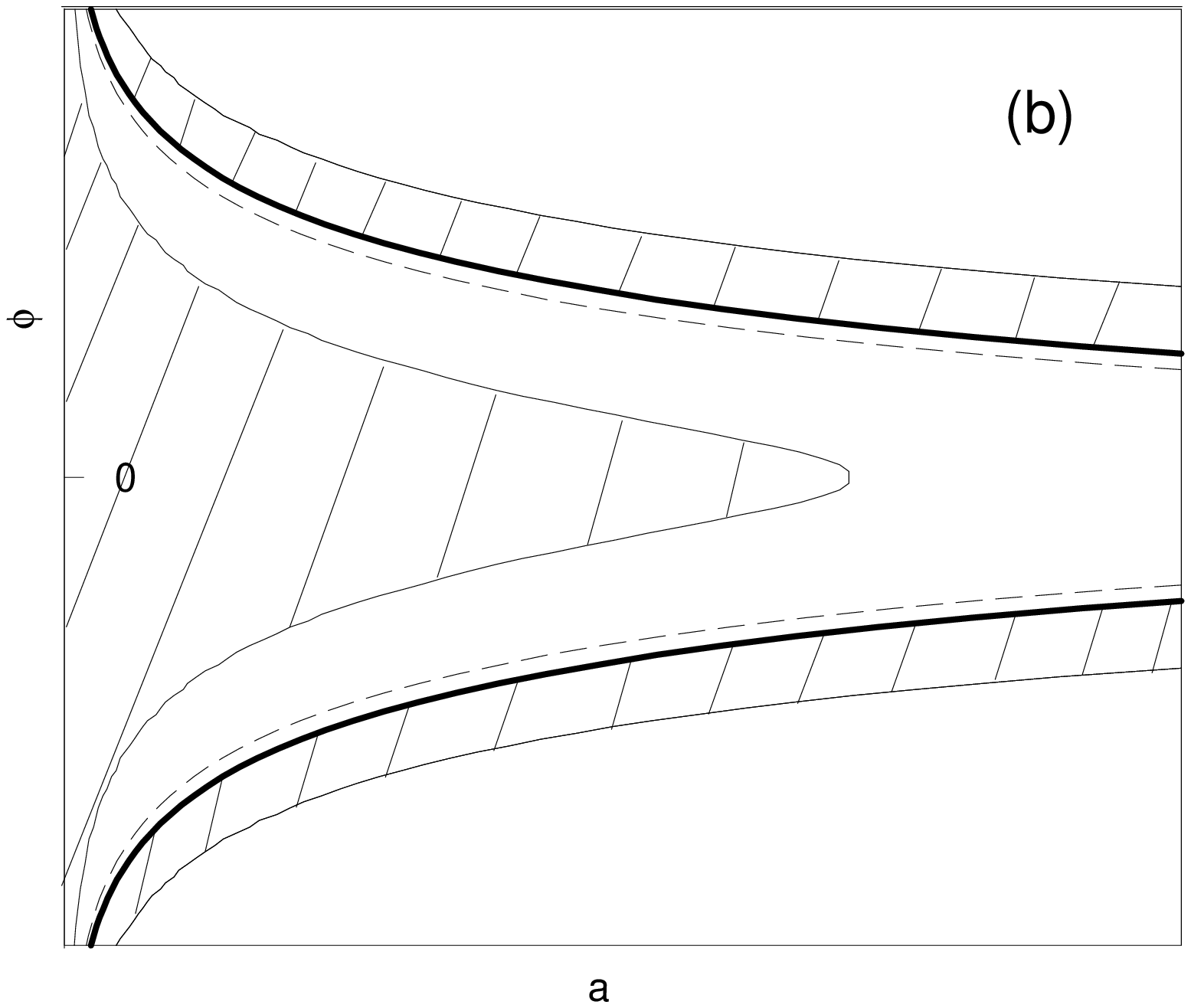}
\includegraphics[scale=0.38, angle=0]{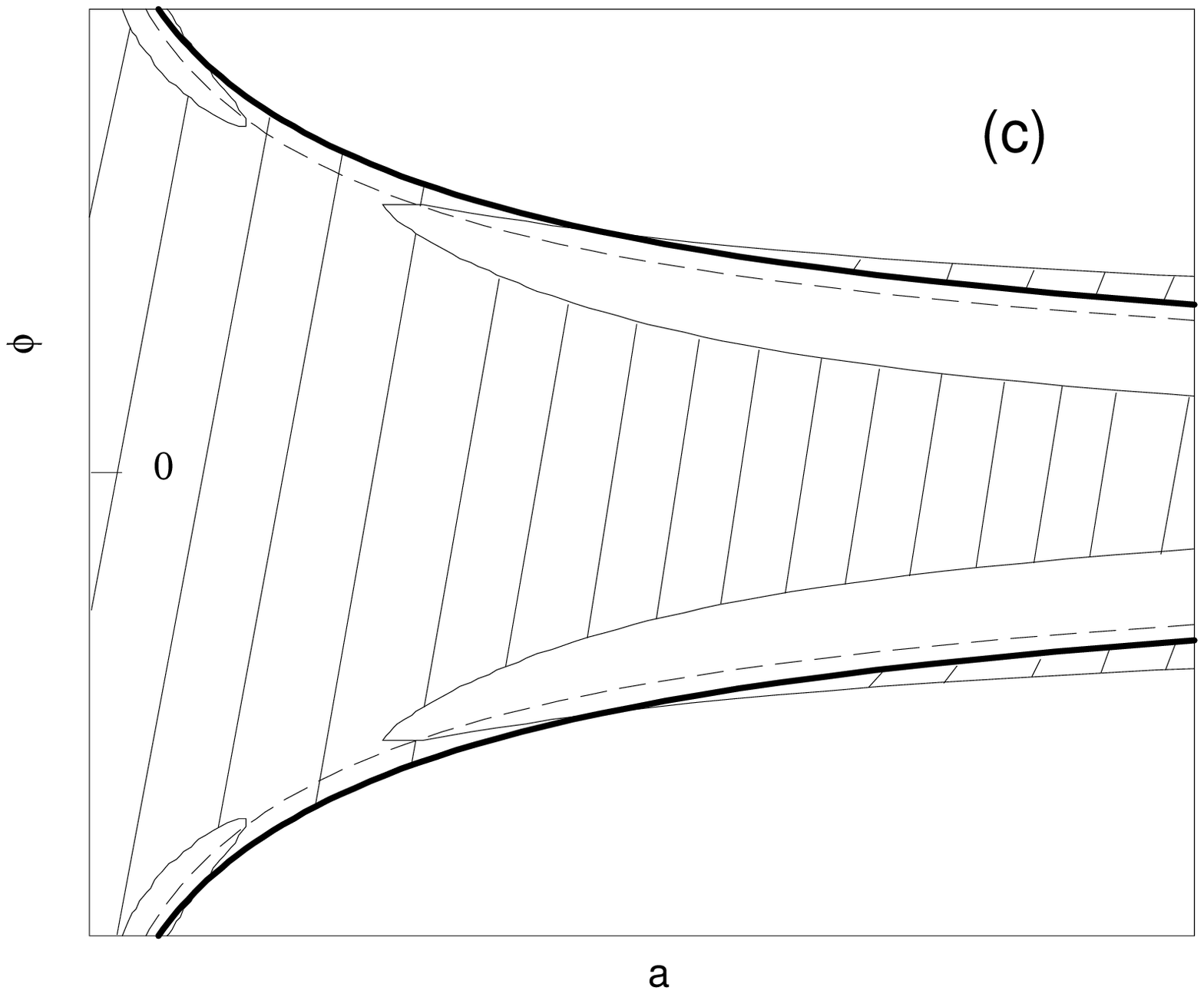}

\caption{Zones of negative curvature (dashed) for the quadratic scalar field potential (a),
for the potential $V=\cosh(\phi)-1$ (b) and 
$V=(exp(\phi/\phi_0)^2+exp(-\phi/\phi_0)^2)-2$ with $\phi_0<0.96 m_P$ (c).
Possible bounces can be located between bold and dashed lines.}
\end{figure}

However, some qualitative difference in configuration of ${\cal R}<0$
 and ${\cal R} >0$ zones exists between the chaotic case in Fig.1(a)
and the regular case in Fig. 1(b). It can be easily checked 
that the curve ${\cal R}=0$ intersects the boundary curve (7) at the Page point (9),
so the picture for the system with Page points (as for power-law
potentials) and without Page points (for steep potentials) should be different.   
Moreover, we can see from the plots that the area of possible 
bounces (which is located between two hiperbolae) belongs to a stable
zone in a regular case, and some part of this area is located in unstable
zone for the chaotic case. This can be checked further for the potential 
$V=(exp(\phi/\phi_0)^2+exp(-\phi/\phi_0)^2)-2$ with $\phi_0<0.96 m_P$, where chaotic
repellor exists for intermediate values of $\phi$ (Fig.1(c)) \cite{we}. All this lead to a suggestion
that negative curvature in zone of bounces is required for the described 
type
of chaotic behavior, though more effort is needed for better understanding this connection.

\section{Islands of regularity}
The main goal of this section is to claim that another type 
(different from described in Introduction) of regular dynamics
exists for the system (3)--(4). This type of dynamics is known for 
a variety of dynamical systems  [15--18]. 
We have found it in the cosmological dynamics ($C=0$) with shallow
potentials. In a general situation this behavior exists even for quadratic potential
if the constant $C$ from the constraint equation (5) is 
sufficiently large and negative \footnote{In nonstandard cosmology this type of dynamics
exists also in braneworlds \cite{resent} and in theories with quadratic curvature corrections
to Einstein gravity\cite{Alexeyev}.}. In all our numerical studies we use the units $m_P/\sqrt{16 \pi}=1$.
\begin{figure}    
\includegraphics[width=0.5\textwidth]{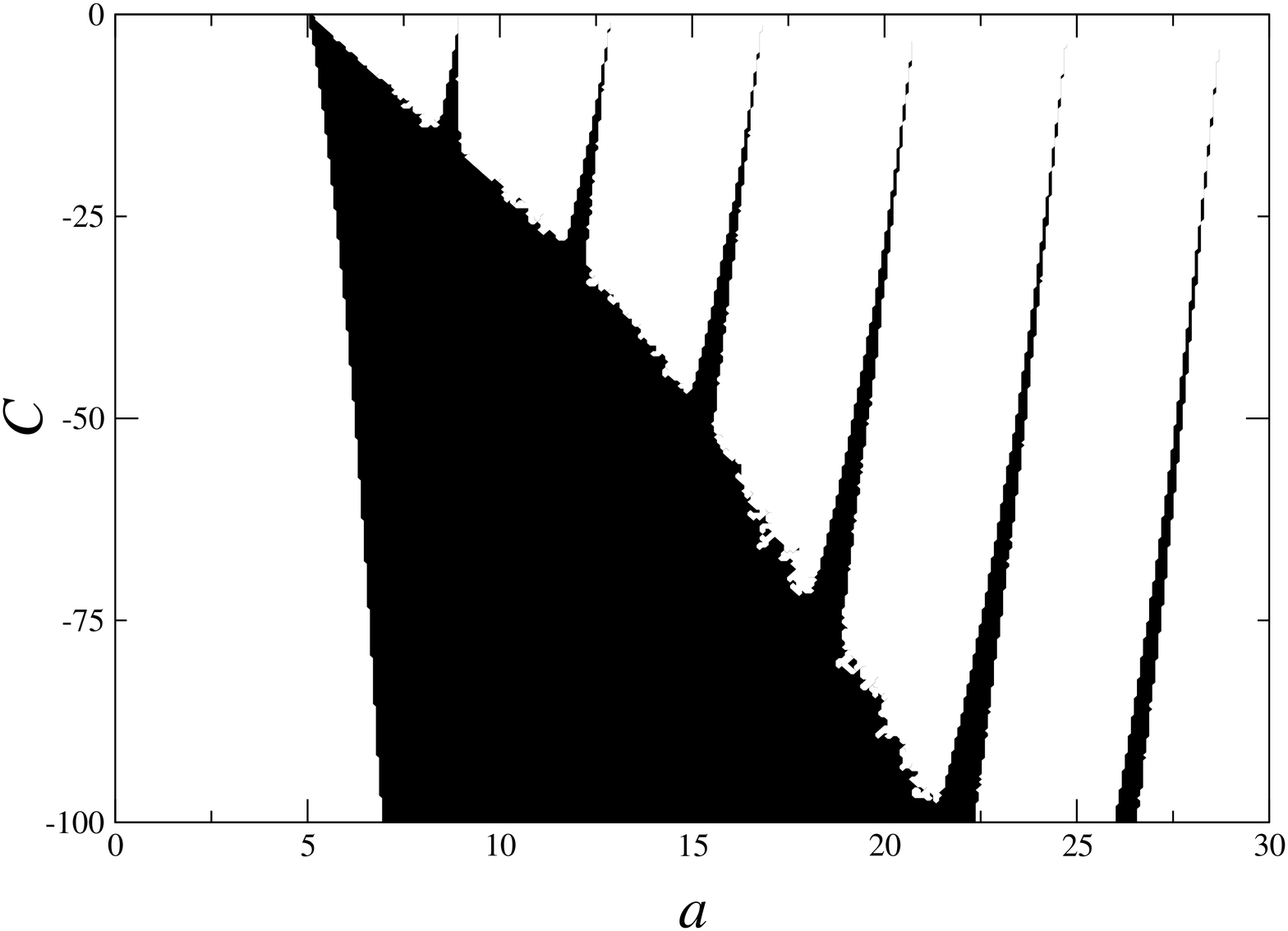}

   \label{f4}
\caption{Zones of initial scale factor for trajectories which have bounce if started
from their points of maximal expansion. For large enough negative $C$ separate intervals of
initial scale factors merge.}
  \end{figure}

It has been already reported that in these two cases the picture of chaotic dynamics changes
significantly in comparison with the cosmological case of power-law potentials.
 In Fig.2 we reproduce the plot from \cite{mysigma}, where zones in initial condition
space leading to at least one bounce is shown
for
the potential $V=m^2 \phi^2/2$ and
negative $C$. A trajectory starts at the point of maximal expansion, so $\dot a=0$,
and using the constraint equation (5) for eliminating $\dot \phi$ we get 2-dimensional space
of initial conditions ($a, \phi$). In  Fig.4 the $\phi=0$ section of this space is
shown depending on $C$.
When some bands, being separate for small $|C|$  merge with $|C|$ increasing,
structure of trajectories changes. Trajectories starting from the wide
zone formed by merged bands show complicated behavior, which can  not be described in a simple
way, like in the $C=0$ case. That is why this type of behavior have been called as "strong" or
"less ordered" chaos. However, the dynamical behavior in this case appeared to be even more
complicated in the sense that islands of regular dynamics can be 
distinguished in a chaotic "sea".

To see this we have studied an appropriate Poincare section, taking for this purpose a point of maximal
expansion, i.e. the point with $\dot a = 0$, $\ddot a < 0$. In Fig. 3 the 
points of this Poincare section are plotted 
for quadratic scalar field potential with $m=1.0$ and $C=-300$. Initial values of the scale factor
in the point of maximal expansion vary from $7$ to $12$.
Island of regular dynamics is easily
distinguishable. It should be pointed out that this regular regime is of completely different nature than
regular dynamics for a steep scalar field potential. Now we can see a quasi-periodic behavior, 
and  trajectories
from this part of the phase space never fall into a singularity. A particular strongly 
stable periodical trajectory
exists for initial $a \sim 8 $  (in general, this value is a function of $C$).

In the vicinity of this trajectory the points of Poincare sections form rather regular ovals,
those from more distant orbits show some structure in their Poincare set. In the transition zone between
regular and chaotic zones  nontrivial structures can appear, like shown in Fig. 4. Here we can see a Poincare
set consisting of 9 disconnected ovals. Structure of zone between chaos and regularity requires further
investigations.

\begin{figure}    
\includegraphics[width=1\textwidth]{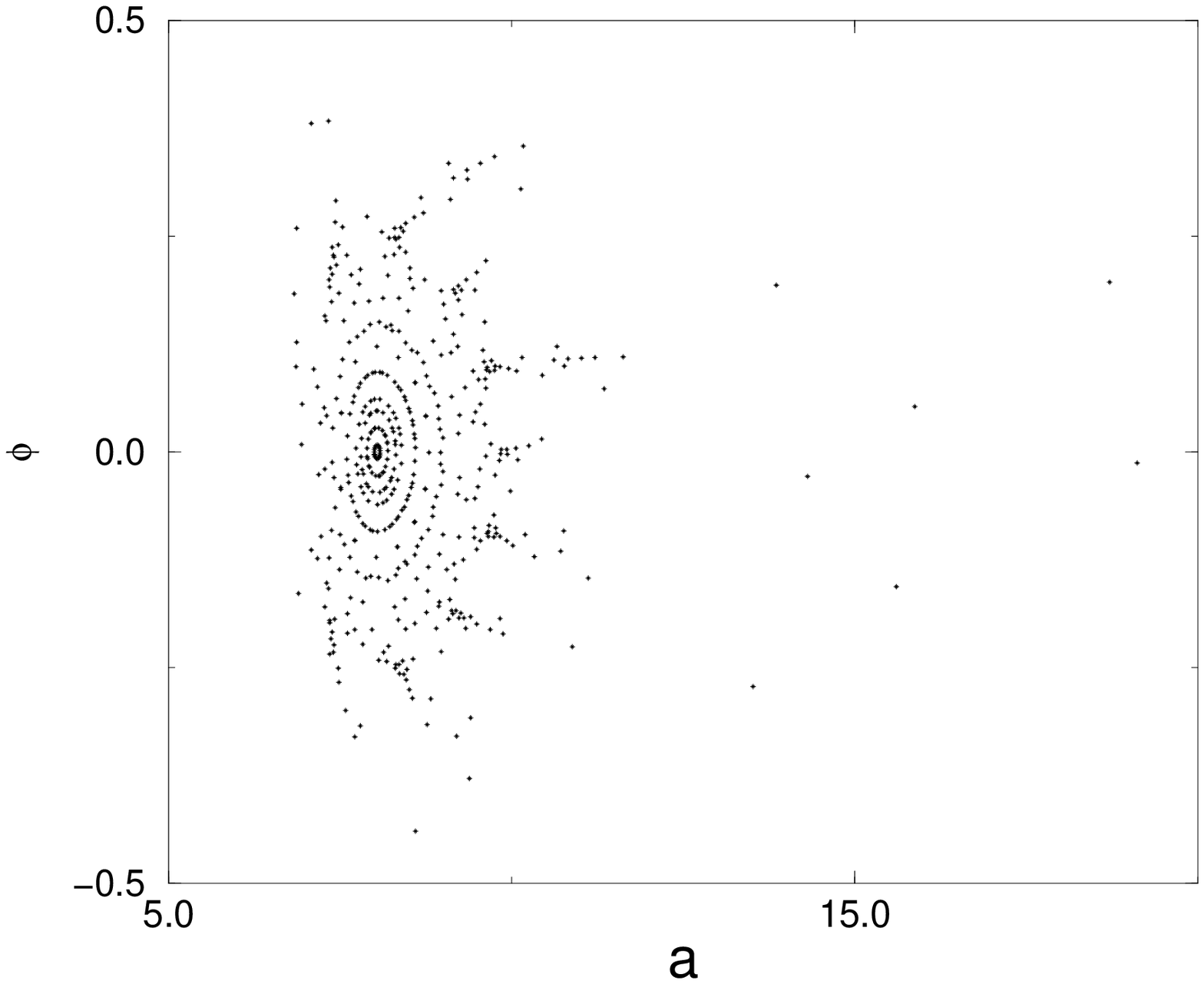}

   \label{f5}
\caption{A Poincare section for trajectories starting in the zone of bounce intervals
merging. The scalar field potential $V(\phi)=m^2 \phi^2/2$, $m=1.0$, the constant $C=-300$.
 Regular structure around initial $a \sim 8, \phi=0$ is clearly seen.}
  \end{figure}

We have seen  similar islands of regularity 
in the cosmological solutions (i. e. with $C=0$) of the system (3)--(4)
with a family of shallow Damour-Mukhanov potentials \cite{Damour} studied 
with respect to their chaotic properties in \cite {our}. 
This family has the form
\begin{equation}
V(\varphi)=\frac{M_{0}^{4}}{q} \left[ \left(1+\frac{\varphi^2}
{\varphi_{0}^{2}} \right)^{q/2}-1 \right].
\end{equation}
with three parameters -- $M_0$, $q$ and $\varphi_0$ and interpolates 
between a quadratic form of the potential for small $\phi$ and a power-law form
$V(\phi) \sim \phi^q$ for large $\phi$.

It was found numerically that the described event of bounce
band merging exists for a subfamily of (13) with $q < 1.24$ and $\phi_0 <
\tilde \phi_0$, where $\tilde \phi_0$ is some function of $q$ \cite{our}. 
In the table below we compare $\tilde \phi_0 (q)$ with the value of $\phi^{'}_0$
for which the island of regularity appears (it exists for $\phi_0<\phi^{'}_0$).
They are close enough, and it 
indicates that bounce zone merging is a good indicator of regularity in our 
system. The difference between these two values grows significantly only in the
vicinity of the critical value of $q \sim 1.24$ when this type of dynamics disappears
for any $\phi_0$.   

\begin{tabular}{|c|c|c|}
\hline
$q$ & $\tilde \phi_0$ & $\phi^{'}_0$ \\
\hline
$0.5$ & $0.5$ & $0.48$ \\
\hline
$1.0$ & $0.25$ & $0.23$ \\
\hline
$1.2$ & $0.06$ & $0.045$ \\
\hline 
$1.21$ & $0.045$ & $0.035$ \\
\hline
$1.22$ & $0.035$ & $0.015$ \\
\hline 
\end{tabular}

 To our knowledge,
this is the first reported case of stable periodical cosmological regime in General Relativity (other
known stable examples like described in \cite{Campos} require modifications 
of Einstein gravity, some other solutions  periodic with respect to scale factor 
evolution in time are 
accompanied by a monotonic growth of a scalar field \cite{Barrow}). This
regular island is surrounded by a "sea" of chaotic trajectories. In general all of them are finite and end in
a cosmological singularity, though existence of chaotic infinite trajectories of non-zero measure is not 
currently ruled out in our numerical studies.

\begin{figure}    
\includegraphics[width=1\textwidth]{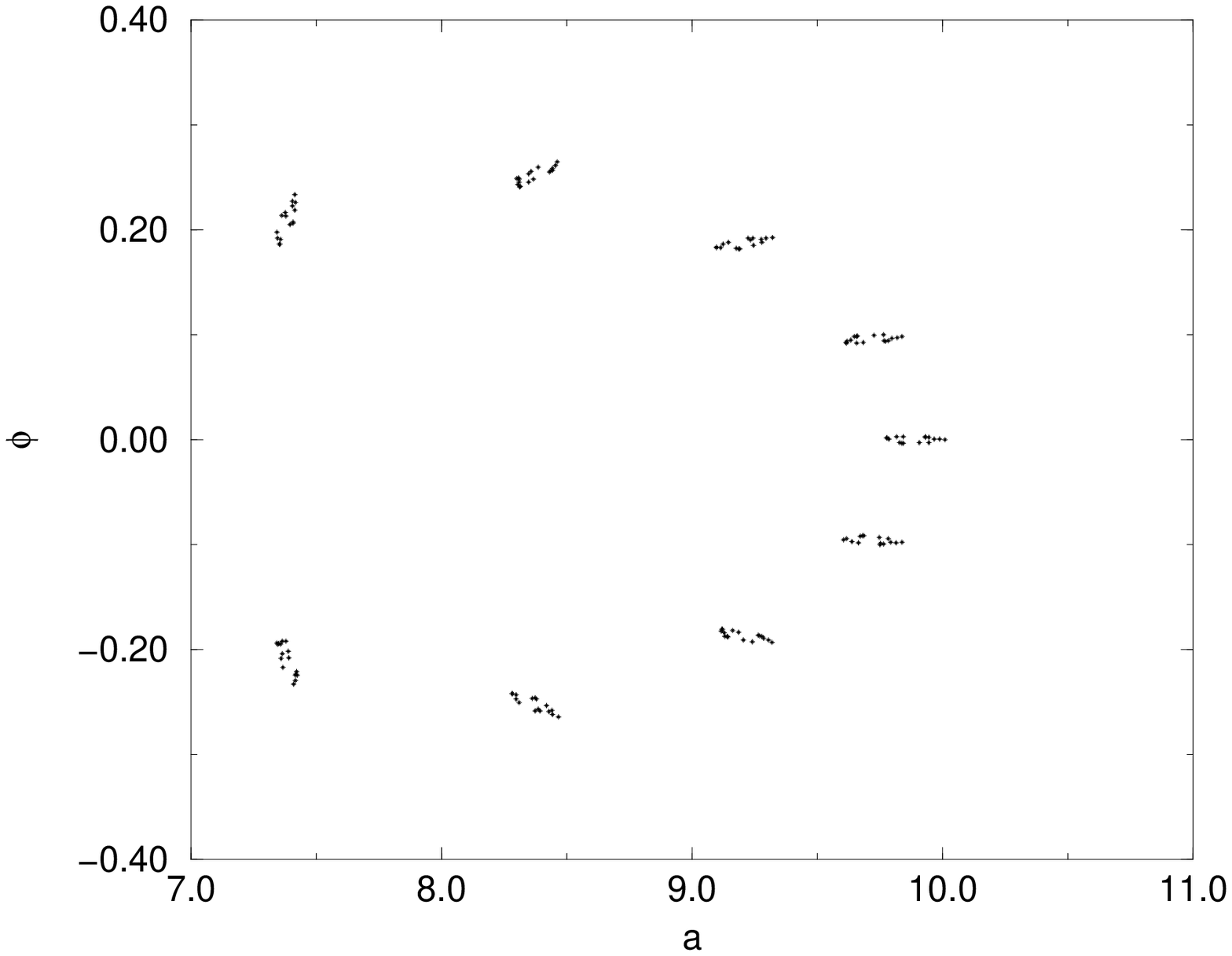}

   \label{f6}
\caption{A Poincare section for a particular trajectory near the boundary of the
island of regularity. The potential and $C$ are the same as in Fig.3.}
  \end{figure}

\section{Conclusions}

We have studied local instability of trajectories representing evolution
of a closed isotropic Universe with a scalar field using the Maupertuis principle. Zones of positive and negative curvature
${\cal R}$ of the configuration space are constructed.
It is shown that the system in question is not a U-system, and there are zones  with positive and negative ${\cal R}$. It is interesting that though  zones with different sign of ${\cal R}$ exist
for both power-law and exponentially steep potential, in the latter case (when the structure of a transient
chaos is absent) the configuration of zones differs qualitatively from the case of less steep potential
(when the chaos is present). Our empirical data suggests that a crucial feature for chaos to exist is local
instability of trajectories in the narrow zones when bounce can occur.

A typical property of dynamical system with nonnegative ${\cal R}$ is that their phase space contains 
positive-measured islands of regular dynamics \cite{Gurzadyan}. We show that the dynamics of this type exists in the case
of sufficiently shallow potential as well as in another similar systems with merged bands of initial conditions leading
to bounces. Trajectories in these islands never reach the cosmological singularity, as well as inflationary regime,
and have trapped for infinite time intervals in a certain zone of small scale factors.

\section*{Acknowledgments}

This work is supported by RFBR grant 05-02-17450 and
scientific school grant 1157.2006.2 of the Russian Ministry
of Science and Technology. A.T. is grateful to Dmitry Anosov, Vahe Gurzadyan,
John Barrow and Jonathan Dawes for discussions.

\end{document}